\documentclass[aps,pre,showpacs,twocolumn,groupedaddress,floatfix]{revtex4}
\usepackage{amssymb,amsmath}
\usepackage{graphicx}
\usepackage{subfigure}
\usepackage[english]{babel}
\usepackage{float}
\usepackage{color}

\begin{document}
\title{Self-trapping transition in nonlinear cubic lattices}
\author{Uta Naether}
\author{Alejandro J. Mart\'inez}
\author{Diego Guzm\'an-Silva}
\author{Mario I. Molina}
\author{Rodrigo A. Vicencio}
\affiliation{Departamento de F\'isica, MSI-Nucleus on Advanced Optics, and Center for Optics and Photonics (CEFOP), Facultad de Ciencias, Universidad de Chile, Santiago, Chile}
\date{\today}
\begin{abstract}
We explore the fundamental question about the critical nonlinearity value needed to dynamically localize energy in discrete nonlinear cubic (Kerr) lattices. We focus on the effective frequency and participation ratio of the profile to determine the transition into localization in one-, two-, and three-dimensional lattices. A simple and general criterion is developed - for the case of an initially localized excitation - to define  the transition region in parameter space (``dynamical tongue'') from a delocalized to a localized profile. We introduce a method for computing the dynamically excited frequencies, which helps us to validate our stationary ansatz approach and the effective frequency concept. A general analytical estimate of the critical nonlinearity is obtained, with an extra parameter to be determined. We found this parameter to be almost constant for two-dimensional systems, and proved its validity by applying it successfully to two-dimensional binary lattices.
\end{abstract}


\maketitle

\section{Introduction}

Discrete nonlinear systems constitute a useful testbed to explore many interesting questions and properties of diverse branches of physics~\cite{pt,rep1,rep2,repBEC,rep3}. Different dimensions and topologies are possible, allowing the description of many different physical phenomena. For instance, in optics, there are recent experimental demonstrations of different nonlinear lattice structures including 3D photonic lattices~\cite{3D,3D2}, 2D ionic-photonic lattices~\cite{ionic}, and also spiraled 3D lattices~\cite{spiral}.
Localized structures are natural solutions in this kind of systems, and they exist when a judicious balance between diffraction and self-focussing effects is reached. At low nonlinearity levels, initial excitations tend to diffract (expand) across the lattice due to the excitation of many  extended states; at larger nonlinearity values, diffraction is inhibited and initial excitations remain localized.
Thus, a very fundamental question related to the necessary conditions to observe a dynamical transition from an extended pattern to a localized one appears.  In other words, how large should the nonlinearity be in order to dynamically localize the excitation? The first attempts to solve this question where focused on small systems of just few sites. In Ref.~\cite{kenkre86} Kenkre and Campbell  studied the self-trapping transition on a nonlinear cubic dimer, finding an exact value for the critical nonlinearity $\gamma_{c}=4$. Then, in Ref.~\cite{mario93} Molina and Tsironis explored systems ranging from 2 to 100 sites. The general conclusion was that as the number of sites increases, the transition approaches the dimer value being, therefore, not an increasing  function of the lattice size. In Ref.~\cite{dunlap93} Dunlap et. al. studied this transition for a nonlinear cubic impurity embedded in a linear lattice and found that the critical nonlinearity was always lower than the one for a full nonlinear system.
Bernstein et al.~\cite{delong93} compared the Hamiltonian of a single-site excitation with the one of a completely homogeneous extended state to find that, for 1D lattices, again, $\gamma_c=4$. In Ref.~\cite{magnus95} Johansson et al. found numerically a lower critical value for 1D lattices ($\gamma_c\approx 3.45$) computed in very large systems and for very long dynamical time scales. This work also explored the transition for 1D binary lattices showing a smaller  value compared to the monoatomic case. These authors mentioned that, by considering simple energy-balance conditions~\cite{delong93}, a rough estimate can be obtained for simple lattices, but not for the binary case. A first attempt to find a more general criterion was done in Ref.~\cite{carlos00} by comparing $\gamma_c$ with the minimum bound-state energy of a nonlinear impurity problem. They found that, for nonlinear  lattices of different dimensions and topologies, there is a kind of universal ratio of $\approx 1.3$ between these two energies. Very recently, Kevrekidis et al.~\cite{kevre08} developed an analytical criterion in terms of Hamiltonian comparisons. This analysis gave a sufficient, but not necessary, condition of $\gamma_c=4$ and $\gamma_c=7.3$ for one- and two-dimensional lattices, respectively.

In the present work we study the problem of dynamical localization in discrete nonlinear cubic (Kerr) lattices of different dimensions and topologies. We develop a simple, but general, criterion to define the regions, in parameter space, where the dynamical transition -- from a completely delocalized profile to a localized one -- occurs. By studying the effective frequency and participation ratio evolution, we are able to identify clearly the localization-delocalization transition and determine numerically the critical value of the nonlinearity. We develop a method to analyze the frequencies excited during the propagation, showing a perfect agreement with the effective frequency approach. We analytically find an expression to predict the necessary nonlinearity to localize energy in any discrete lattice, obtaining a lower estimate when compared to all previous analytical estimations. By doing this, an extra unknown parameter appears which is determined numerically from our data. We observe that for two-dimensional lattices, this parameter approaches a constant value, and that its size decreases with the dimension of the problem. Finally, we explore the validity of this constant value by studying 2D binary (ionic or diatomic)~\cite{ionic} cubic lattices.

\section{Model}

The propagation of waves in nonlinear cubic lattices is well described by a discrete nonlinear Schr\"odinger (DNLS) equation~\cite{pt,rep1,rep2,repBEC}:
%
\begin{equation}
-i\frac{d u_{\vec{n}}}{d z}=\epsilon_{\vec{n}} u_{\vec{n}} + \sum_{\vec{m}\neq \vec{n}}u_{\vec{m}}+\gamma|u_{\vec{n}}|^2 u_{\vec{n}}\ ,
\label{dnls}
\end{equation}
%
where $u_{\vec{n}}$ corresponds to the wave amplitude at site $\vec{n}$ in a D-dimensional lattice of $N$ sites. The coupling between different lattice sites is restricted to the nearest-neighbors of site $\vec {n}$. [For example, for 1D lattices the coupling interaction is given by $(u_{n+1}+u_{n-1})$, while for 2D-rectangular ones become $(u_{n+1,m}+u_{n-1,m}+u_{n,m+1}+u_{n,m-1})$]. Parameter $\epsilon_{\vec{n}}$ corresponds to the energy (propagation constant) at the ${\vec{n}}$-th site. $z$ corresponds to the dynamical coordinate~\cite{pt,rep1,rep2,repBEC}. We will vary the nonlinear coefficient $\gamma$ in order to study the necessary amount of nonlinearity to transit from a delocalized to a localized profile. Model (\ref{dnls}) possesses two conserved quantities, the Norm 
\[
Q\equiv\sum_{\vec{n}} |u_{\vec{n}}|^2,
\]
and the Hamiltonian
\[
H\equiv\sum_{\vec{n}} \left[\epsilon_{\vec{n}} |u_{\vec{n}}|^2+\left(\sum_{\vec{m}}u_{\vec{m}}u_{\vec{n}}^*+c.c.\right)+(\gamma/2) |u_{\vec{n}}|^4\right].
\]
Extended linear ($\gamma=0$) stationary solutions of model (\ref{dnls}) exist inside the linear spectrum ($\{ \lambda_{bottom},\lambda_{top}\}$ for constant $\epsilon_{\vec{n}}$). Along this work, we will focus on the following lattices: 1D ($\lambda_{top}=2$), 2D-Honeycomb (2D-h, $\lambda_{top}=3$), 2D-square (2D-s, $\lambda_{top}=4$), 2D-triangular (2D-t, $\lambda_{top}=6$), and 3D-square (3D, $\lambda_{top}=6$). 

We will study the self-trapping transition by considering a single-site (delta-like) excitation: $u_{\vec{n}}(z=0)=\delta_{\vec{n},\vec{n}_0}$, where $\vec{n_0}$ corresponds to a position well inside the lattice (i.e., a bulk excitation for which the reflection from boundaries is negligible). Considering that $Q$ and $H$ are dynamical constants, this input condition implies fixed values $Q_0=1$ and $H_0=\epsilon_{\vec{n}_0}+\gamma/2$. (A variation of $\gamma$ is equivalent to changing $Q$, the key parameter when thinking on experimental realizations~\cite{exp1,exp2,exp3,exp4,exp5,exp6,exp7}).
We use the participation ratio, defined as $R\equiv Q^2/\sum_{\vec{n}}|u_{\vec{n}}|^4$
as an indicator of the degree of localization of a wave-packet (e.g., $R=1$ for a single-site excitation, and $R=N$ for an equally excited array). 

\subsection{Effective frequency}

To study the dynamical evolution and its relation with the stationary solutions of the system, we will consider an approximation introduced in Ref.~\cite{predisorder}. This approach assumes every instantaneous profile as a set of different, linear and nonlinear, stationary modes characterized by a single instantaneous effective frequency. We consider a given profile as a stationary solution of the form $u_{\vec{n}}(z)=u_{\vec{n}} \exp{(i \lambda_e z)}$, being $\lambda_e$ the \textit{effective or average frequency} of the excited modes on this profile. By replacing this in (\ref{dnls}), multiplying by $u_{\vec{n}}^*$ and summing in all the sites along the lattice, we get an expression similar to the Hamiltonian.
We rewrite this expression in terms of the fundamental quantities and parameters of model (\ref{dnls}), obtaining that $\lambda_e Q=H+\gamma Q^2/(2R)$.
However, $Q=Q_0$ and $H=H_0$, therefore we obtain a closed and simple expression for the effective frequency:
\begin{equation}
\lambda_e =\epsilon_{\vec{n}_0}+\frac{\gamma}{2}\left(1+\frac{1}{R}\right).
\label{lam1}
\end{equation}
Before the
self-trapping transition, the wave-packet diffracts/delocalizes over the whole lattice and $R(z_{max})\sim N\gg 1\Rightarrow \lambda_e (z_{max}) \approx \epsilon_{\vec{n}_0}+\gamma/2$ [$z_{max}$ depends on the particular dimension, size, and topology of the lattice]. On the other hand, after the self-trapping transition, localization dominates and $R(z_{max})\sim 1$, implying that $\lambda_e (z_{max})\approx \epsilon_{\vec{n}_0}+\gamma$. Therefore, for nonlinear cubic lattices, we have a region in parameter-space -- \textit{dynamical tongue} -- inside which all the dynamics is contained: $\gamma/2 \leqslant \lambda_e- \epsilon_{\vec{n}_0} \leqslant \gamma$.

\section{Numerical results}
%
\begin{figure}[h]
\centering
\includegraphics[width=8.5cm]{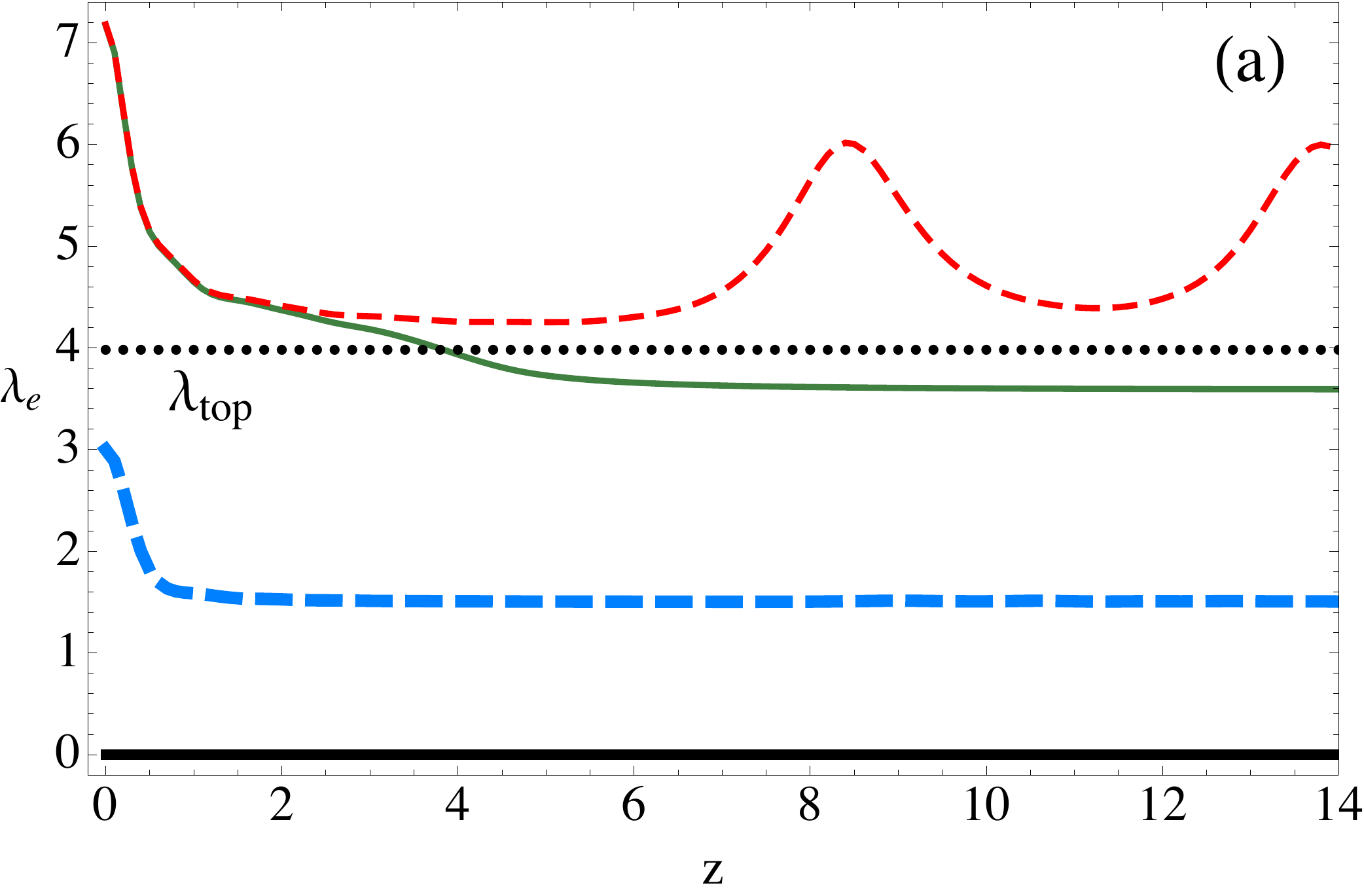}\\
\includegraphics[width=8.5cm]{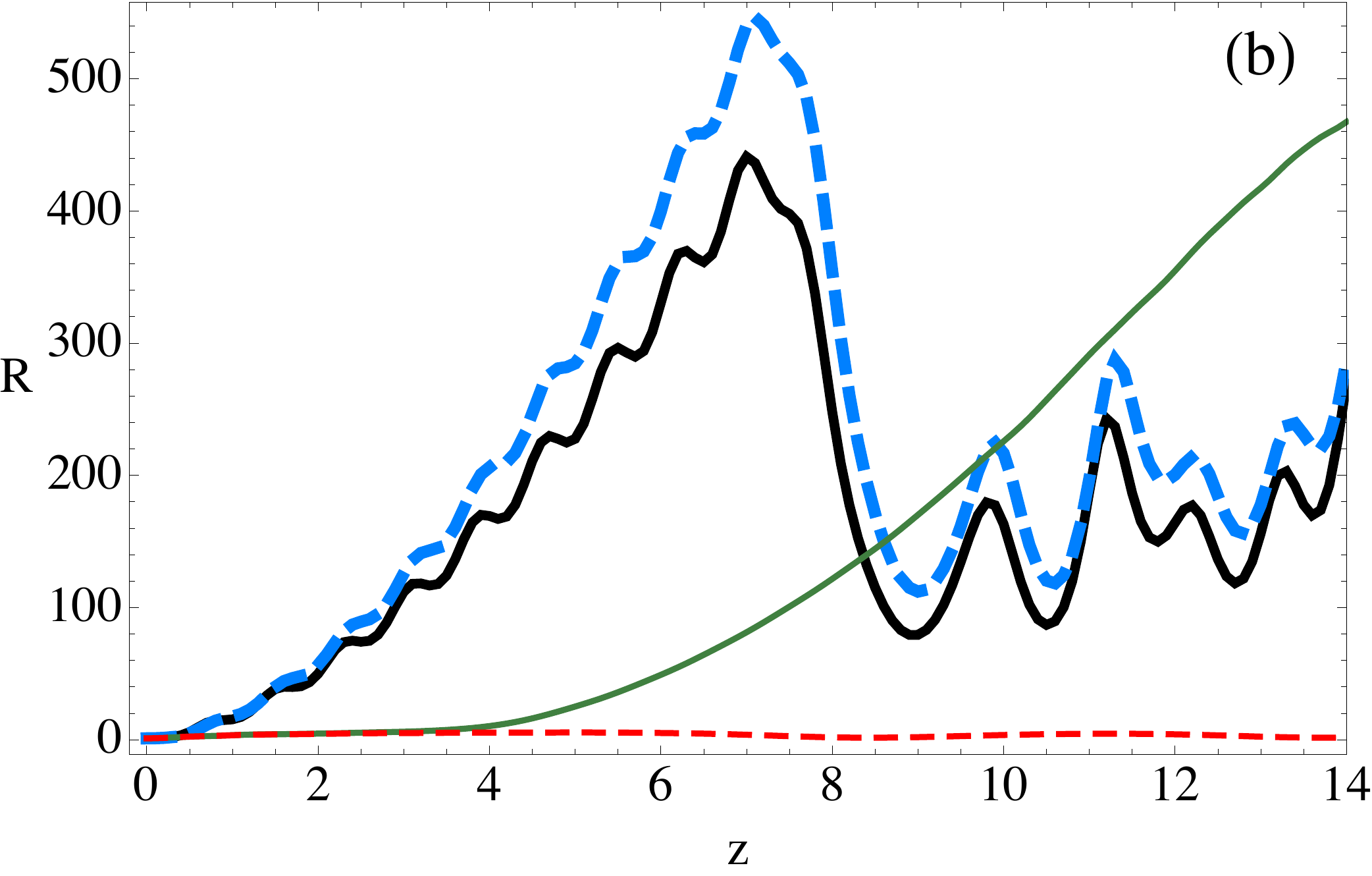}
\includegraphics[width=2.85cm]{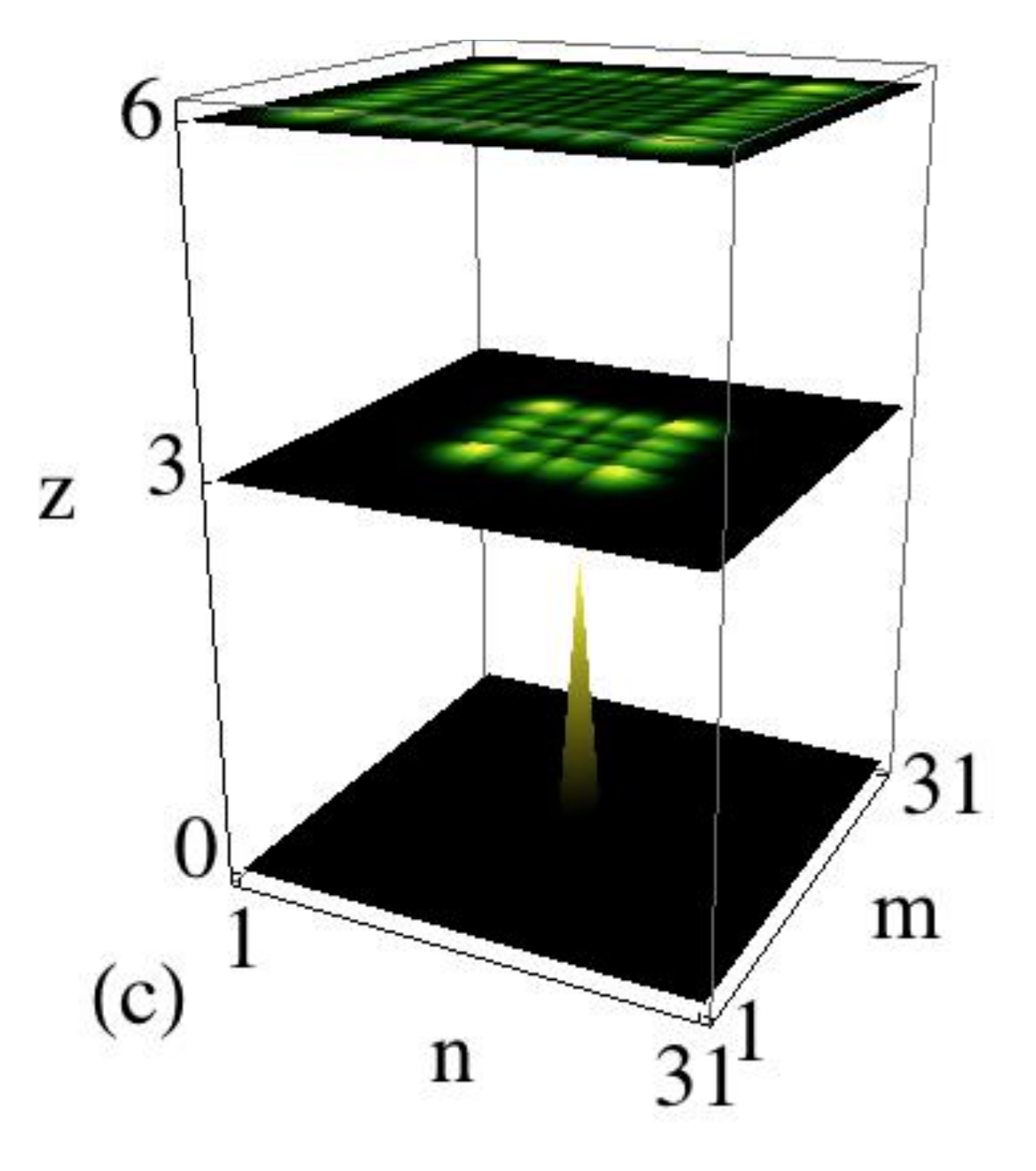}
\includegraphics[width=2.85cm]{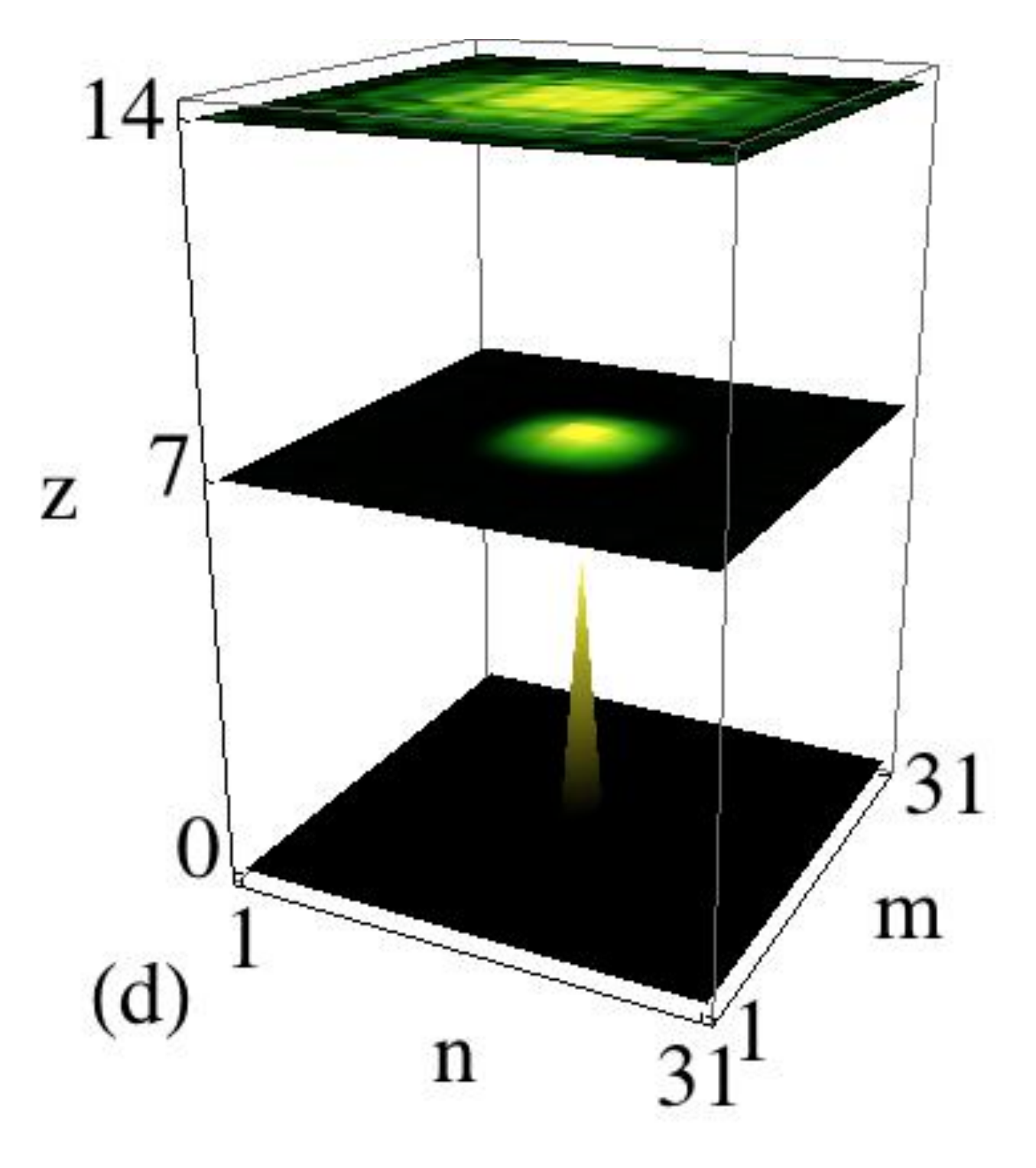}
\includegraphics[width=2.85cm]{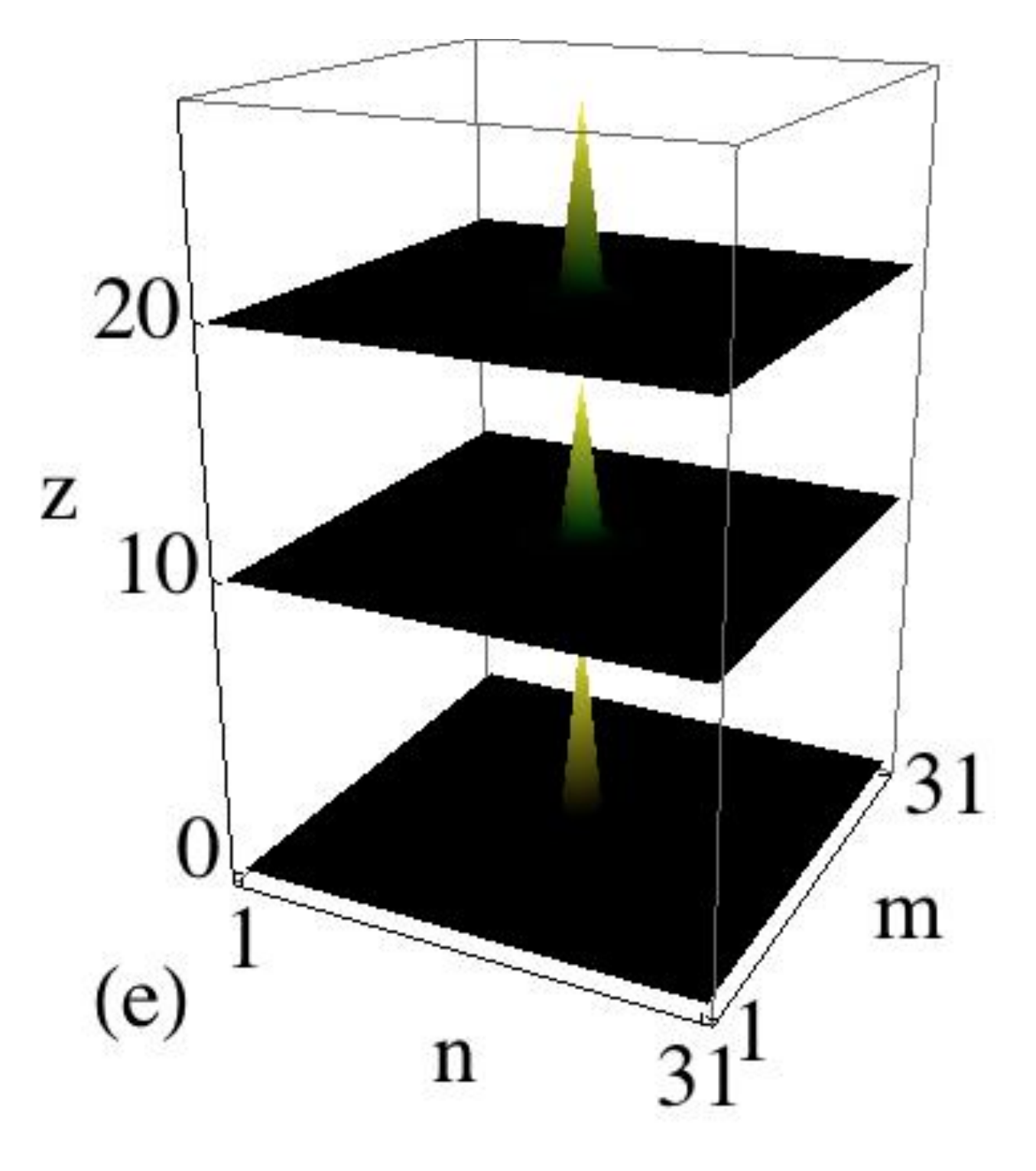}
\caption{(Color online) Example for a 2D-square lattice: (a) $\lambda_e$ and (b) $R$ versus $z$ for $\gamma =0$, $3$, $7.17$, and $7.18$ represented by black (thick-full), blue (thick-dashed), green (thin-full), and red (thin-dashed) lines, respectively. Dynamical evolution ($|u_{n,m} (z)|^2$) examples for (c) $\gamma=3$, (d) $\gamma=7.17$, and (e) $\gamma=7.18$. $N=31\times 31$ sites and $\epsilon_{\vec{n}_0}=0$.}
\label{fig1}
\end{figure}
%
By differentiating (\ref{lam1}) with respect to $z$ we obtain  $\partial \lambda_e/ \partial z =-(\gamma/2R^2) \partial R/\partial z$. Therefore, below the transition we must observe (on average) an evolution with an increasing $R$ (diffusion) and a decreasing $\lambda_e$, if $\gamma>0$. Figure~\ref{fig1} shows an example for a 2D-square lattice where we observe that, before the self-trapping transition [black (thick-full), blue (thick-dashed), and green (thin-full) curves], the effective frequency decreases from $\gamma$ to $\approx \gamma/2$ and that $R$ increases from $R=1$ to $R\sim N\gg 1$. For small values of $\gamma$ [black (thick-full) and blue (thick-dashed) curves], the wave-packet approaches the lattice boundary at $z_{max}$, where $R$ is a maximum (see in Fig.~\ref{fig1}(b) that for $z>z_{max}\sim 7$, $R$ decreases due to the reflections at the boundaries). Since $\lambda_e$ is not a conserved quantity of model (\ref{dnls}), its evolution will show some oscillations around an average value [not observed in the scale of Fig.~\ref{fig1}(a)], implying that different, linear and nonlinear, frequencies are being excited. When the transition into localization takes place [red (thin-dashed) curves] the final average frequency becomes larger than $\gamma/2$, and stays outside of the linear band ($\lambda_e>\lambda_{top}$). The profile starts to excite more nonlinear frequencies reducing the excitation of linear modes and, as a consequence, it becomes localized (a similar process occurs in disordered lattices when studying nonlinear delocalization transitions~\cite{predisorder}).  For the  2D-square example, self-trapping is observed for $\gamma\geqslant\gamma_c=7.18$. The oscillation in $\lambda_e$ implies an oscillation (of opposite sign) of the $R$-value [see red (thin-dashed) curves in Figs.~\ref{fig1}(a) and (b)]. Figs.\ref{fig1}(c)-(e) show some dynamical propagation examples for different values of $\gamma$.

\subsection{Effective versus average frequency}
%
\begin{figure}[h]
\centering
\includegraphics[width=8.23cm]{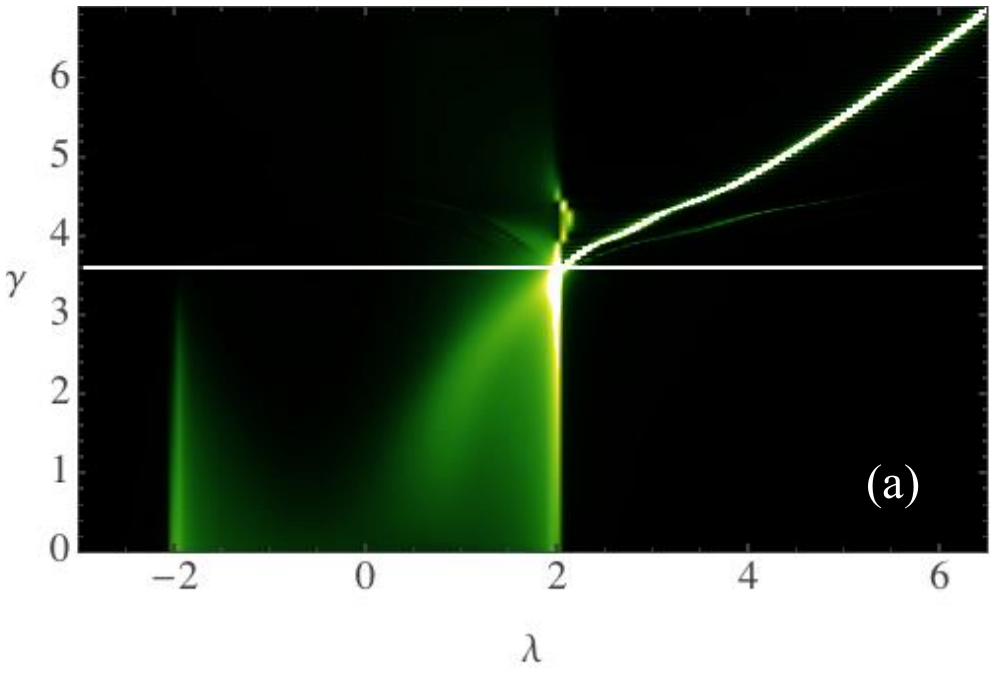}
\includegraphics[width=8.23cm]{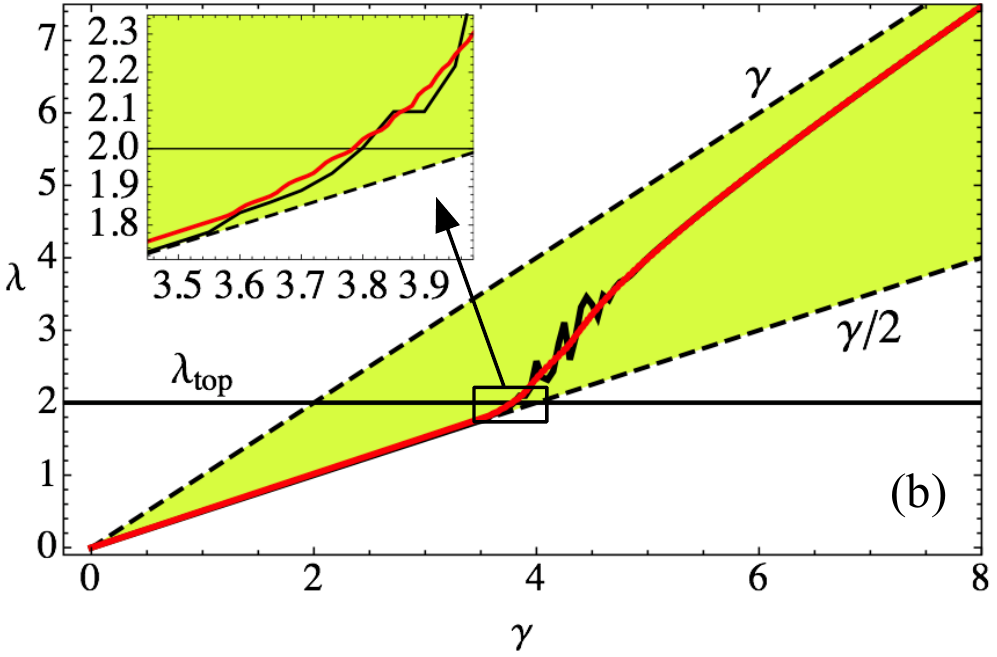}
\caption{(Color online) Example for a 1D lattice with $N=401$: (a) Spectral density versus $\gamma$ and $\lambda$. (b) $\lambda_e (z_{max})$ (black) and $\left<\lambda \right>$ [red (gray)] versus $\gamma$, inside the dynamical tongue (shaded area).}
\label{fig0}
\end{figure}
%
We interpret $\lambda_e$ as an average frequency of the profile that includes all the modes excited during the propagation. To validate this interpretation, we integrate system \eqref{dnls} directly and compute the longitudinal Fourier transform of the amplitude evolution in each waveguide, $\tilde{u}_n(\lambda)\equiv \int dz\ u_n(z)\exp(- 2\pi i \lambda z)$, to obtain the normalized spectral density
\begin{equation}
 g(\lambda,\gamma)=\sum_n |\tilde{u}_n(\lambda)|^2/\sum_{n,\lambda}|\tilde{u}_n(\lambda)|^2. 
\end{equation} 
Figure \ref{fig0}(a) shows this quantity for a 1D lattice. We see how a positive nonlinearity modifies the original distribution of frequencies, exciting more modes closer to $\lambda_{top}=2$. For $\gamma\gtrsim 3.6$ [see the horizontal line in Fig.~\ref{fig0}(a)], an emerging nonlinear mode is strongly excited (brightest peak) and self-trapping starts to occur. In addition, we calculate a mean frequency, defined as $\left<\lambda \right>=\sum_{\lambda}\lambda\ g(\lambda,\gamma)$ and plot, in Figure~\ref{fig0}(b), a comparison between this quantity and $\lambda_e$. We see how both curves deviate from the $\gamma/2$-line in almost the same region ($\gamma\gtrsim 3.6$) [see Fig.~\ref{fig0}(b)-inset]. In general, there is a very good agreement between both quantities. Only above the threshold, there is some disagreement due to the fluctuations in $\lambda_e$. For higher dimensional lattices, we found a better agreement because fluctuations are reduced. These results validate our assumption of $\lambda_e$ as an effective and average frequency and thus, useful when studying different dynamical processes~\cite{predisorder}.

\subsection{Dynamical tongue and self-trapping transition}

\begin{figure}[h]
\centering
\includegraphics[width=8.31cm]{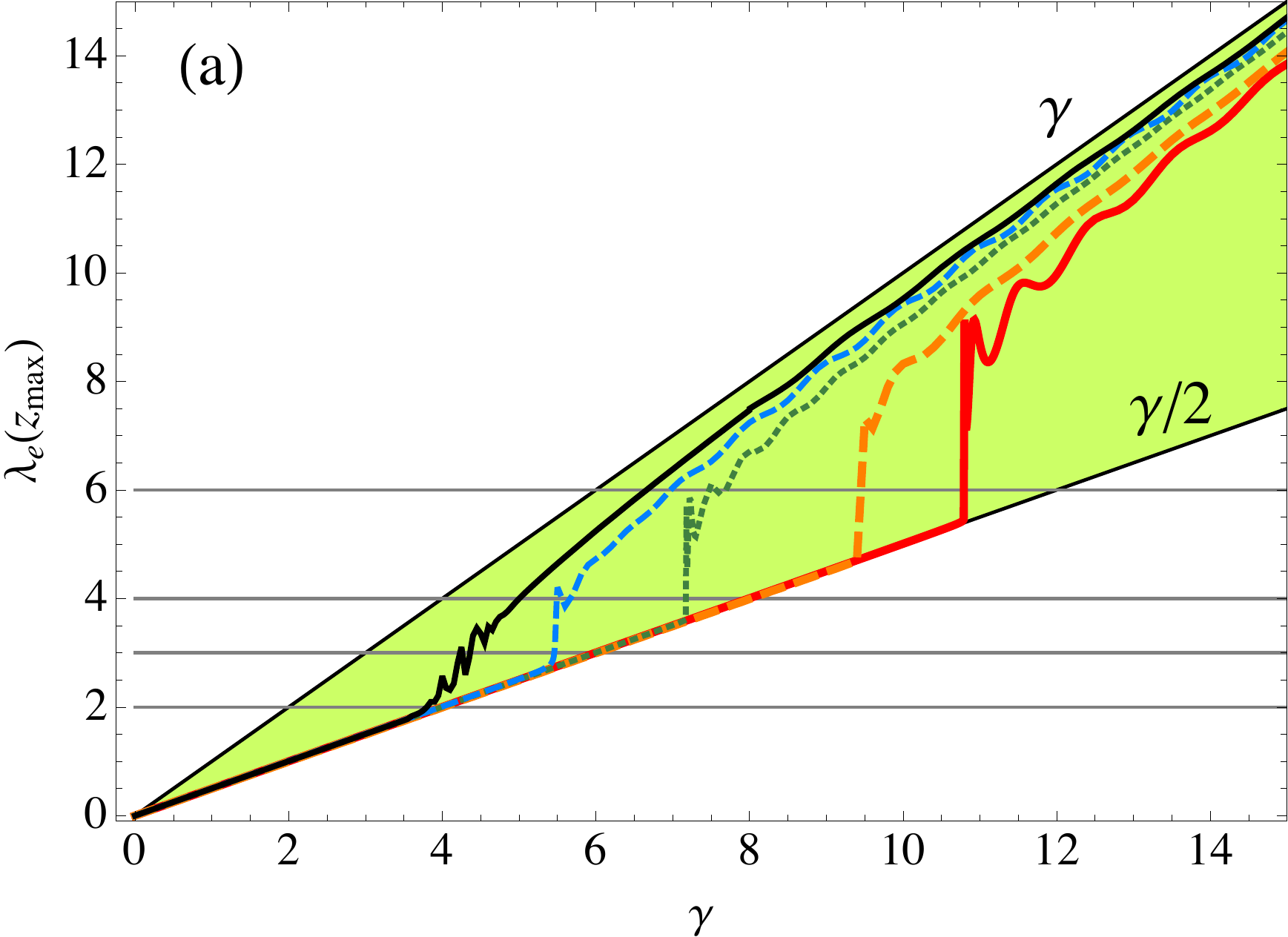}\vspace{-0.3cm}
\includegraphics[width=8.31cm]{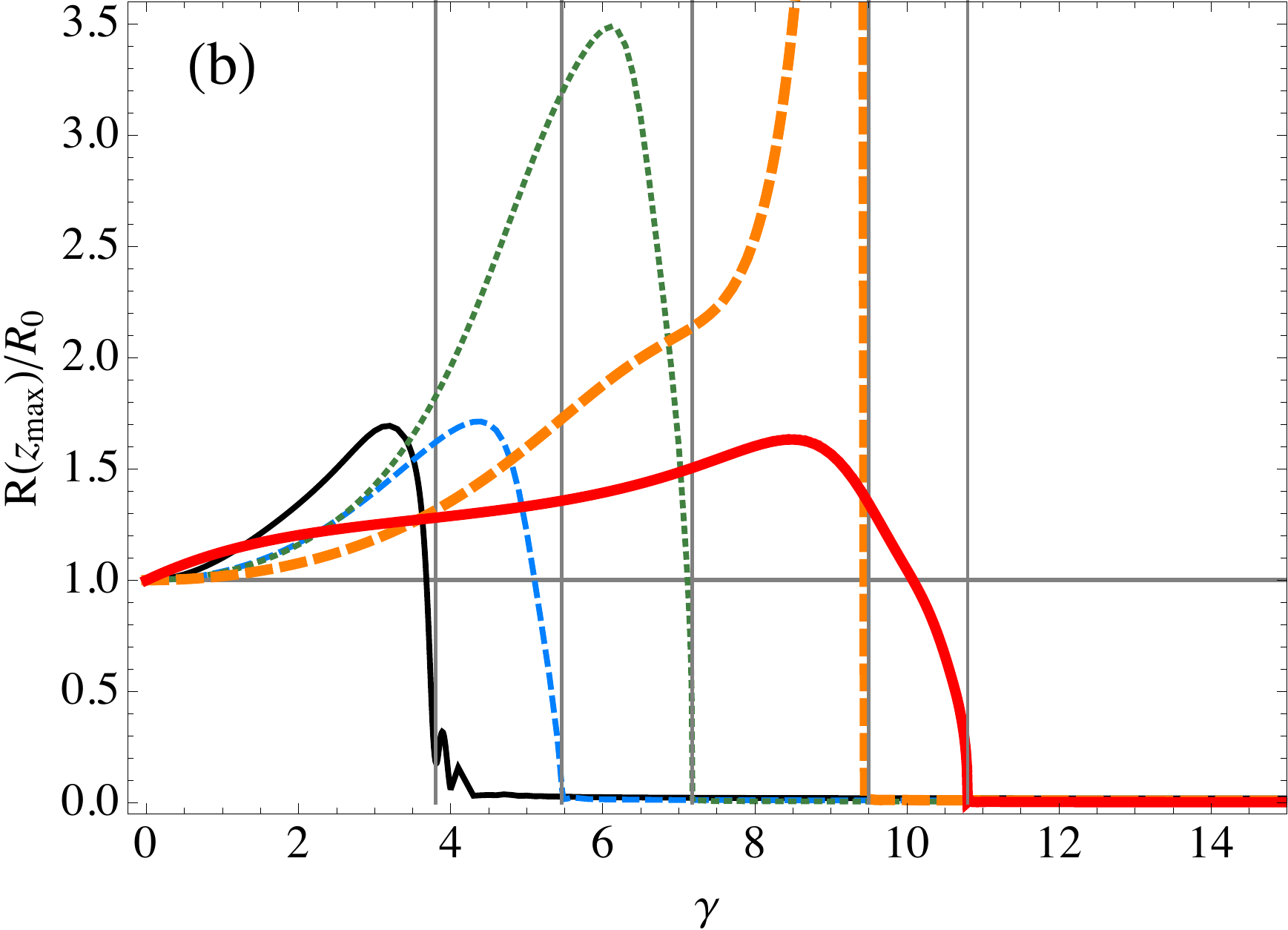}
\caption{(Color online) (a) $\lambda_e(z_{max})$ and (b) $R(z_{max})/R_0$ versus $\gamma$ for 1D [black (thin-full)], 2D-h [blue (thin-dashed)], 2D-s [green (thin-dotted)], 2D-t [red (thick-full)], and 3D [orange (thick-dashed)]. Horizontal lines in (a) represent $\lambda_{top}$. Vertical lines in (b) denote $\gamma_c$. For all lattices $\epsilon_{\vec{n}_0}=0$.}
\label{fig2}
\end{figure}

In Fig.~\ref{fig2}, we collect our numerical results for different cubic lattices. Fig.~\ref{fig2}(a) shows that, for each lattice, all the dynamics is contained inside the ``dynamical tongue'' (shaded area), in agreement with our analytical prediction. When nonlinearity is small, the effective frequency is $\approx \gamma/2$ because the wave-packet have diffracted [Figs.~\ref{fig1}(c) and (d)]. When the self-trapping occurs [see Fig.~\ref{fig1}(e)], $\lambda_{e}$ increases in the direction of $\gamma$. For most of the studied lattices, we observe that the transition implies a rather abrupt jump in frequency ($\partial \lambda_e/\partial \gamma\gg 1$). For one-dimensional systems, the transition is softer due to the absence of power thresholds to create a nonlinear localized mode~\cite{flach,kalo}. The excited nonlinear modes emerge from the top of the band [see Fig.~\ref{fig0}(a)] and the interaction with linear modes is stronger (in fact, for $\gamma\sim 3.6$, the average frequency is smaller than $\lambda_{top}$ implying that there is a mixture of linear and nonlinear excited modes). For dimension $D\ge2$ there is a threshold to excite nonlinear modes~\cite{flach}, causing a fast jump in $\lambda_e$ above the band edge (the interaction of linear and nonlinear modes is weaker and the transition becomes sharper).
%
%

Fig.~\ref{fig2}(b) summarizes the results for the participation ratio normalized to $R_0\equiv R(\gamma=0,z_{max})$, which corresponds roughly to the size of the diffracted profile in the linear regime. We see that the self-trapping transition implies a fast change in the participation ratio ($\partial R/\partial \gamma\ll -1$) from values of the order of the lattice size to values of the order of few lattice sites. This figure shows that $R$ is a very useful parameter to trace the self-trapping transition, by showing the effective number of excited sites. This figure also shows that the transition is always abrupt, and its sharpness increases with dimension. By collecting the data from Fig.~\ref{fig2}, we obtain the approximate numerical $\gamma_c$-values shown in Table~\ref{tabla}.
We define each one as the value of $\gamma$ for which $\lambda_{top}=\lambda_e(z_{max})$. These values were obtained for fixed lattice sizes and fixed propagation distances, However, for different parameters we obtain essentially the same $\gamma_c$-values. Our aim is to obtain a robust numerical estimation for this transition. We  should keep in mind that these transitions may occur and be observed in real physical systems, where an estimation range is more relevant~\cite{pt,rep1,rep2,repBEC}). In fact, if we use another quantity as an indicator (for instance, the space-averaged fraction of power~\cite{carlos00}), we observe the occurrence of the self-trapping transition in similar regions.
%
\begin{table}[h]
\caption{\label{tab:table1}%
Numerical results of $\gamma_c$ and $R_c$.}
\begin{ruledtabular}
\begin{tabular}{cccccccc}
\textrm{Lattice}&\textrm{1D}&\textrm{2D-h}&\textrm{2D-s}&\textrm{2D-t}&\textrm{3D}\\
\colrule
$\gamma_c$ &3.8  & 5.46 & 7.18  &10.8 & 9.5\\
$R_c$ &19 & 10.1 & 8.8 & 9 & 3.8 \\
\end{tabular}
\end{ruledtabular}
\label{tabla}
\end{table}

\section{Analytical estimates}

Now, we tackle the non-trivial problem of getting an analytical estimation of the \textit{critical nonlinearity} ($\gamma_c$), for which the self-trapping transition occurs. In general, previous analysis observed that, at least, $|\gamma_c|\sim|\lambda_{top}-\lambda_{bottom}|$ (which depends strongly on the particular dimension and topology of the lattice). This criterion essentially says that when the nonlinear contribution ($\gamma Q$) is larger than the linear one (approximately, the size of the band), the localization tendency will be more important than the diffractive one, and the wave-packet will tend to localize. This rough criterion predicts $\gamma_c$ values larger than the actual numerical ones~\cite{magnus95,kevre08}. Figures~\ref{fig1} and \ref{fig2} show that the transition implies an abrupt change of the size of the wave-packet (determined by $R$) with a final effective frequency [$\lambda(z_{max})$] moving out of the band. Our hypothesis is quite simple: If the effective  frequency is inside the band, the profile interacts with more linear modes (which are extended) tending to delocalization. However, if the effective frequency is outside the linear band, the wave-packet excites a set of nonlinear frequencies and only few linear modes, tending to localization. In general, the transition will depend on the particular linear properties of the lattice but, also, on the sign of the nonlinearity. In this work, we focus only in the case $\gamma>0$, and on lattices possessing symmetric linear bands with respect to $\lambda=0$ (i.e., $|\lambda_{top}|=|\lambda_{bottom}|$). In this case, the $|\gamma_c|$-value is independent on the sign of the nonlinearity and the data from Table~\ref{tabla} also applies for $\gamma<0$. (For non symmetric linear bands, our method also applies being $\gamma_c$ different depending on the sign of the nonlinearity). Thus, for $\gamma>0$, we conjecture that the \textit{critical nonlinearity} is obtained when the effective frequency (\ref{lam1}) coincides with the border of the linear band; i.e., when
\begin{equation}
\lambda_{e}=\lambda_{top}\Rightarrow\gamma_{c}=2 \left(\lambda_{top}-\epsilon_{\vec{n}_0}\right) \left(\frac{R_c}{1+R_c}\right)\ .
\label{gc}
\end{equation}
In other words, we will be able to excite a nonlinear localized state when the average frequency stays outside of the band and resonances with linear modes are reduced or cancelled. We see that equation (\ref{gc}) predicts a lower value of the critical nonlinearity than previous estimates. However, we get an extra parameter defined as $R_c$. We conjecture that this parameter corresponds to a critical size of the wave-packet for which self-trapping starts to occur, a kind of minimum volume for a profile to be considered as localized. In Table~\ref{tabla}, we present our computation of $R_c$ for all the studied lattices using Eq.~(\ref{gc}) and the numerically obtained $\gamma_c$-values. From this data, we see that the critical effective size decreases as the dimension increases, which is in agreement with the localization tendency of localized states in nonlinear cubic lattices. Close to the band, 1D nonlinear stationary solutions are very broad, while for 2D lattices, and even more for 3D ones, nonlinear stationary solutions are very localized above the norm threshold. For 2D lattices, we numerically found that this value is around $10$, leading to an estimation for $\gamma_c$ which is $\approx 90\%$ of the previous analytical predictions. It is interesting to notice that $R_c$ is nearly constant for two-dimensional discrete nonlinear cubic lattices, which are the most explored lattices nowadays in different contexts of experimental physics~\cite{pt,rep1,rep2,repBEC,rep3}.

\section{Binary 2D-square lattices}

Finally, we want to go further and explore the validity and robustness of our findings in more complex settings, particularly in binary lattices~\cite{ionic}. We focus on a 2D-square array with site-energies $\epsilon_{2\vec{n}+1}=0$ and $\epsilon_{\vec{2n}}=\Delta \epsilon$, being $\Delta \epsilon$ the site-energy contrast. In this case, the location of the input excitation plays a fundamental role in the determination of the transition. The factor ``$\lambda_{top}-\epsilon_{\vec{n}_0}$'' in expression (\ref{gc}) implies a lower critical value for a larger $\epsilon_{\vec{n}_0}$. After a straightforward calculation, we obtain the band-edge frequency $\lambda_{top}=0.5(\Delta \epsilon+\sqrt{64+\Delta \epsilon^2})$. Considering the critical effective size as constant for 2D lattices ($R_c=10$), we get the following estimation for the critical nonlinearity
\begin{equation}
\gamma_{c}\approx0.9 \left(\sqrt{64+\Delta \epsilon^2}+\Delta \epsilon-2\epsilon_{\vec{n}_0}\right) .
\label{gcbina}
\end{equation}
%
%
\begin{figure}[t]
\centering
\includegraphics[width=8.5cm]{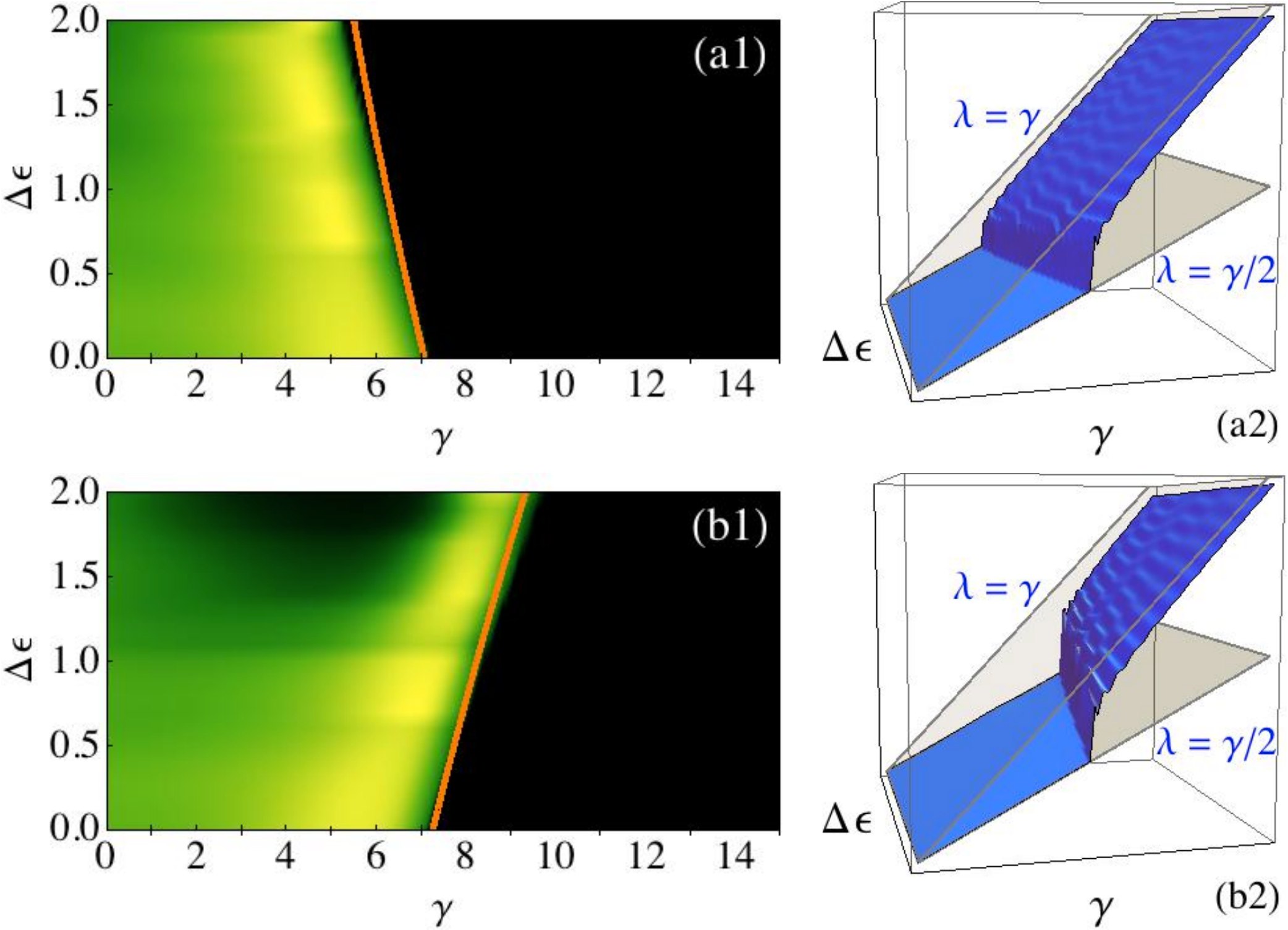}
\caption{(Color online) Results for a binary 2D-square lattice. (a1) [(b1)] shows $R$ versus $\{\gamma,\Delta\epsilon\}$ for $\epsilon_{\vec{n}_0}=\Delta \epsilon$ [$\epsilon_{\vec{n}_0}=0$]. (a2) [(b2)] shows $\lambda_e (z_{max})-\epsilon_{\vec{n}_0}$ versus $\{\gamma,\Delta\epsilon\}$ for $\epsilon_{\vec{n}_0}=\Delta \epsilon$ [$\epsilon_{\vec{n}_0}=0$]. $N=31\times 31$ sites.}
\label{fig3}
\end{figure}
%
Therefore, for $\epsilon_{\vec{n}_0}=\Delta \epsilon$ we expect a reduction of $\gamma_c$ if the contrast increases ($\Delta \epsilon\gg1\Rightarrow \gamma_c\rightarrow0$). On the other hand, for $\epsilon_{\vec{n}_0}=0$ we expect an increment of the critical nonlinearity if the contrast increases. Figure \ref{fig3} collects our findings for these cases. The density plots (a1) and (b1) show the participation ratio versus $\gamma$ and $\Delta \epsilon$; dark (bright) color means an smaller (larger) $R$-value, and the estimate (\ref{gcbina}) is plotted by orange straight lines. We see an almost perfect agreement between the numerics and the analytical estimation, validating the use of a constant $R_c$-value for 2D lattices. Figures (a2) and (b2) show the transition in terms of the effective frequency. We clearly see the opposite tendencies depending on the input site-energy. The quantity $\lambda_e(z_{max})-\epsilon_{\vec{n}_0}$ is effectively $\gamma/2$ below the transition (bright flat surface), increasing in the direction of $\gamma$ above $\gamma_c$. The transition in participation ratio and frequency is very abrupt and described 
with great accuracy by (\ref{gcbina}).


\section{Conclusions}

In conclusion, we have studied the problem of the self-trapping transition for different nonlinear cubic lattices, giving a rather general panorama of this very fundamental issue. We showed theoretically and numerically that, by considering a single-site input excitation, all the dynamics is contained in a very precise parameter-region in the shape of a ``dynamical tongue''. We have found approximate numerical values for the critical nonlinearity where this transition occurs. We showed that the effective frequency can be understood as an average quantity, giving a good insight of the frequencies participating in the dynamics. We developed analytically a new formula to predict the self-trapping transition, that makes use of an extra parameter (effective critical size) that depends strongly on the dimension of the system, being smaller for higher dimensions. Binary lattices are well described by our analytical prediction when combined with the obtained effective size value.

\begin{acknowledgements}
The authors wish to thank M. Johansson and S. Rojas-Rojas for useful discussions. This work was supported in part by FONDECYT Grants 1110142, 1120123, a CONICYT doctoral fellowship, Programa ICM P10-030-F, and Programa de Financiamiento Basal de CONICYT (FB0824/2008).\\
\end{acknowledgements}


\begin{thebibliography}{}

\bibitem{pt}D.K. Campbell, S. Flach, and Y.S. Kivshar, Phys. Today \textbf{571}, 43 (2004).

\bibitem{rep1}S. Flach and A. Gorbach, Phys. Rep. {\bf 467}, 1 (2008).

\bibitem{rep2}F. Lederer, G.I. Stegeman, D.N. Christodoulides, G. Assanto, M. Segev, and Y. Silberberg, Phys. Rep. \textbf{463}, 1 (2008).

\bibitem{repBEC}O. Morsh and M. Oberthaler, Rev. Mod. Phys. {\bf 78}, 179 (2006).

\bibitem{rep3} Z. Chen, M. Segev, and D. N. Christodoulides, Rep. Prog. Phys. {\bf 75}, 086401 (2012).

\bibitem{3D}P. Zhang, R. Egger, and Z. Chen, Opt. Express {\bf 17}, 13151 (2009).

\bibitem{3D2}P. Zhang, N.K. Efremidis, A. Miller, Y. Hu, and Z. Chen, Opt. Lett. {\bf 35}, 3252 (2010). 

\bibitem{ionic}P. Zhang, S. Liu, C. Lou, F. Xiao, X. Wang, J. Zhao, J. Xu, and Z. Chen, Phys. Rev. A {\bf 81}, 041801 (2010).

\bibitem{spiral}J. Xavier, S. Vyas, P. Senthilkumaran, C. Denz, and J, Joseph,  Opt. Lett. {\bf 36}, 3512 (2011).

\bibitem{kenkre86}V.M. Kenkre and D.K. Campbell, \prb \ \textbf{34}, 4959 (1986).
\bibitem{mario93}M.I. Molina and G.P. Tsironis, Physica D \textbf{65}, 267 (1993).

\bibitem{dunlap93}D.H. Dunlap, V.M. Kenkre, and P. Reineker, \prb \ \textbf{47}, 14842 (1993).

\bibitem{delong93}L.J. Bernstein, K.W. DeLong, and N. Finlayson, Phys.Lett.A \textbf{181}, 135 (1993).

\bibitem{magnus95}M. Johansson, M. H\"ornquist, and R. Riklund, \prb \textbf{52}, 231 (1995).

\bibitem{carlos00}C.A. Bustamante and M.I. Molina, \prb \textbf{62}, 15287 (2000).
\bibitem{kevre08}P.G. Kevrekidis, J.A. Espinola-Rocha, Y. Drossino, and A. Stefanov, Phys.Lett.A \textbf{372}, 2247 (2008).

\bibitem{exp1}H.S. Eisenberg, Y. Silberberg, R. Morandotti, A.R. Boyd, J.S. Aitchison, \prl \textbf{81}, 3383 (1998).

\bibitem{exp2}F. Chen, M. Stepi\'c, C. R\"uter, D. Runde, D. Kip, V. Shandarov, O. Manela, and M. Segev, Opt. Exp. \textbf{13}, 4314 (2005).

\bibitem{exp3}A. Szameit, D. Bl\"omer, J. Burghoff, T. Schreiber, T. Pertsch, S. Nolte, A. T\"unnermann, and F. Lederer, Opt. Exp. \textbf{13}, 10552 (2005).

\bibitem{exp4}D. Neshev, E. Ostrovskaya, Y. Kivshar, and W. Krolikowski, \ol \ \textbf{28}, 710 (2003).

\bibitem{exp5}Z. Chen, H. Martin, E. Eugenieva, J. Xu, and J. Yang, Opt. Exp. \textbf{13}, 1816-1826 (2005).

\bibitem{exp6}T. Anker, M. Albiez, R. Gati, S. Hunsmann, B. Eiermann, A. Trombettoni and M.K. Oberthaler, \prl \ \textbf{94}, 020403 (2005).

\bibitem{exp7}I. Bloch, Nature Physics \textbf{1}, 23 (2005).

\bibitem{predisorder}R.A. Vicencio and S. Flach, \pre \ \textbf{79}, 016217 (2009).

\bibitem{flach} S. Flach, K. Kladko, and R.S. MacKay, \prl \ \textbf{78}, 1207 (1997).

\bibitem{kalo} G. Kalosakas, K.\O. Rasmussen, and A.R. Bishop, \prl \ \textbf{89}, 030402 (2002).


%
%
%
%
%
%
%
\end{thebibliography}
\end{document}